\documentclass[lettersize,journal]{IEEEtran}
\usepackage{amsmath, amssymb, amsfonts}
\usepackage{algorithmicx}
\usepackage{algorithm}
\usepackage[noend]{algpseudocode}
\usepackage{algorithmicx, algorithm}
\usepackage{diagbox}
\usepackage{multirow}
\usepackage{graphicx}
\usepackage{textcomp}
\usepackage{xcolor}
\usepackage{subfig}
\usepackage{array}
\usepackage{verbatim}
\usepackage{threeparttable}
\usepackage{tabularx}
\usepackage{booktabs}
\usepackage{threeparttable}
\usepackage{makecell}
\usepackage[unicode=true]{hyperref}
\usepackage{epstopdf}
\usepackage{epsfig}
\usepackage{tikz} 
\usepackage{amsfonts}
\usepackage{amsmath}

\usepackage{color}

\usepackage{enumitem}

\usepackage{algorithm}  
\usepackage{algpseudocode}  
\usepackage{amsmath}

\usepackage{alphalph} 

\ifCLASSOPTIONcompsoc
  \usepackage[nocompress]{cite}
\else
  \usepackage{cite}
\fi

\ifCLASSINFOpdf

\else

\fi

\begin{document}

\title{Affiliation-based Local Community Detection across Multiple Networks}

\author{Li Ni, Zhou Xie,  Yiwen Zhang*,   Wenjian Luo, Senior Member, IEEE, and Victor S. Sheng, Senior Member, IEEE
\IEEEcompsocitemizethanks{
\IEEEcompsocthanksitem Li Ni, Zhou Xie, and  Yiwen Zhang are with the School of Computer Science and Technology, Anhui University, Hefei, Anhui, 230601, China.
Li Ni is also with Guangdong Provincial Key Laboratory of Novel Security Intelligence Technologies, Shenzhen, 518055, China.

Wenjian~Luo is with the School of Computer Science and Technology, Harbin Institute of Technology, Shenzhen, 518055, China.

Victor S.Sheng is with the  Department of Computer Science, Texas Tech University, Lubbock,  TX 79409 USA.

Email: nili@ahu.edu.cn,  e22201043@stu.ahu.edu.cn,  zhangyiwen@ahu.edu.cn, luowenjian@hit.edu.cn, and victor.sheng@ttu.edu. (Corresponding author: Yiwen Zhang)

}}

\markboth{Journal of \LaTeX\ Class Files,~Vol.~14, No.~8, August~2021}%
{Shell \MakeLowercase{\textit{et al.}}: A Sample Article Using IEEEtran.cls for IEEE Journals}


\maketitle

\begin{abstract}
Real-world networks are often constructed from different sources or domains, including various types of entities and diverse relationships between networks, thus forming multi-domain networks. A single network typically fails to capture the complete graph structure and the diverse relationships among multiple networks. Consequently, leveraging multiple networks is crucial for a comprehensive detection of community structures. Most existing local community detection methods discover community structures by integrating information from different views on multi-view networks. However, methods designed for multi-view networks are not suitable for multi-domain networks. Therefore, to mine communities from multiple networks, we propose a Local Algorithm for Multiple networks with node Affiliation, called LAMA, which is suitable for both multi-view and multi-domain networks. The core idea of LAMA is to optimize node affiliations by maximizing the quality of communities within each network while ensuring consistency in community structures across multiple networks. The algorithm iteratively optimizes node affiliations and expands the community outward based on affiliations to detect the community containing the seed node. Experimental results show that LAMA outperforms comparison algorithms on two synthetic datasets and five real datasets.
\end{abstract}

\section{Introduction}
\IEEEPARstart{C}{ommunity} detection is a fundamental task in the analysis of complex networks. Unlike global community detection algorithms that analyze the entire network, local community detection rapidly identifies a specific community using only localized data\cite{cite2_5581103}, thus garnering attention from researchers.
Numerous local community detection algorithms have been developed, including methods based on local modularity \cite{cite4_article,cite25_ni2019community}, dynamic affiliation functions \cite{cite44_luo2018local}, higher-order graph clustering \cite{cite23_yin2017local}, and semi-supervised learning \cite{cite52_DBLP:journals/tkde/NiGZLS24,cite63_10.1145/3394486.3403154}.
These methods identify community structure from a single network. 
Real-world networks often exhibit associations and information from a single network is incomplete \cite{cite64_10.1145/3336191.3371806,cite49_ni2018co}.

\begin{figure}[!t]
    \centering
    \includegraphics[width=0.5\textwidth]{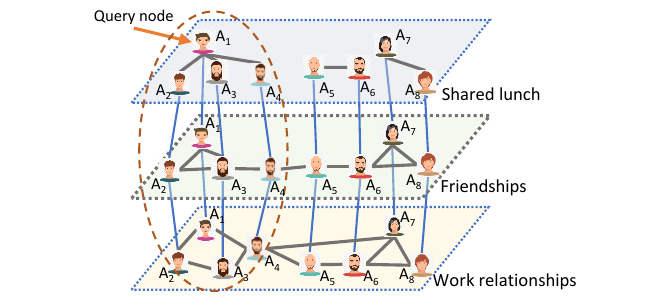}
    \caption{A example of multi-view networks}
    \label{multi-view networks}

    \vspace{0.3cm}

    \includegraphics[width=0.5\textwidth]{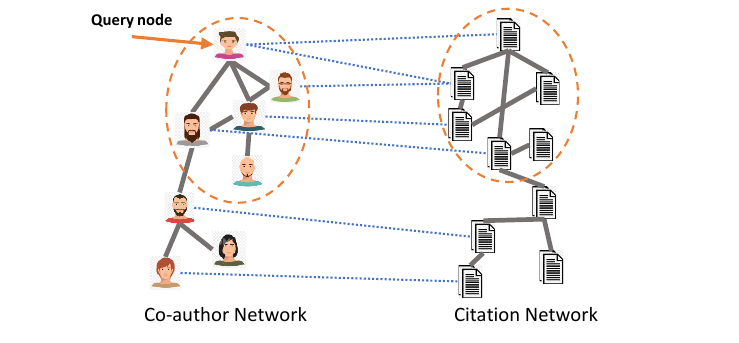}
    \caption{A example of multi-domain networks}
    \label{multi-domain networks}
\end{figure}

Therefore, utilizing information from multiple networks to mine community structures has attracted much attention \cite{cite60_9395530,cite61_DBLP:journals/bigdatama/YangW18}. 
Luo et al.\cite{cite6_10.1145/3394486.3403069} categorize multiple networks into two types: 1) Multi-view networks. 
It refers to representations of the same set of nodes from different views, where each network's edges depict distinct types of relationships. 
Fig. \ref{multi-view networks} demonstrates the relationship among three different aspects for eight employees: (a) shared lunch, (b) friendships, and (c) work relationships\cite{cite48_DBLP:journals/sigmod/KimL15}.
2) Multi-domain multi-networks. It involves different types of nodes and various relationships that may exist between them \cite{cite49_ni2018co}.
Fig. \ref{multi-domain networks} illustrates a co-author network, where nodes are authors and edges represent collaborations, and a citation network, where nodes are papers and edges denote citations between them.
For multi-view networks, Li et al. \cite{cite46_li2018community} identify the core nodes and subsequently extend the local community containing the given seed through local random walks originating from the core nodes. 
Pournoor et al. \cite{cite27_pournoor2021propagation} consider both intra-layer and inter-layer probabilities to navigate towards the most probable neighbors, and compute node scores to identify the local community.
For multi-domain networks, Luo et al. \cite{cite6_10.1145/3394486.3403069} compute the transition probabilities of nodes in networks and combine them with conductance to identify communities containing the seed node. These transition probabilities account for both the intra-layer probability within the current network and the transition probabilities from other networks.

Existing studies designed for multi-view networks are not applicable for mining communities from multi-domain networks \cite{cite46_li2018community,cite27_pournoor2021propagation}.
Although the work in \cite{cite6_10.1145/3394486.3403069} is suitable for multi-domain networks, it overlooks the consistency of communities across networks and requires access to the entire network. 
However, community structures across networks often exhibit consistency. For example, scholars with shared research interests commonly form author communities through collaborative publications. 
Similarly, research papers often cite papers similar to their topic, thus building a paper community around a specific research topic.
Consequently,  scholars who publish papers in the same paper community tend to have the same research interest, demonstrating a high level of thematic consistency. 
Therefore, we propose the problem of Local Community Detection in Multiple Networks (LCDMN),  aimed at maximizing the quality of community structures within each network while ensuring consistency across networks, utilizing only local information.
To address this problem, we develop an approach, called LAMA, which is suitable for mining community structures in both multi-view and multi-domain networks. 
LAMA gradually expands the community outward from seed nodes, simultaneously maintaining the quality of the community within each network and the consistency of the community across networks during the expansion process. 
Our contributions are summarized as follows:

\begin{itemize}
    \item The existing method on multi-domain networks traverses the entire network to detect a community that contains the seed node, which requires access to more network information than a local algorithm. Besides, it ignores the consistency of community between networks. Therefore, we propose the LCDMN problem, aiming at maximizing the quality of communities within each network while ensuring consistency in community structures across multiple networks by only visiting local network around the seed node.
    \item To address the LCDMN problem, we propose a novel algorithm built on node affiliation, called LAMA, with only local information. It iteratively optimizes node affiliations and expands local communities based on node affiliations. Furthermore, LAMA ensures the consistency of communities across different networks and adaptively determines different weights for each network.
    \item    Experiments are conducted on five real-world and two synthetic datasets. 
Experimental results show that LAMA yields better communities than that detected by three multi-view local community detection algorithms and one multi-domain community detection algorithm. 
\end{itemize}

The remainder of the paper is organized as follows. 
Section \ref{Section II Problem Statement} introduces the problem statement. 
The proposed algorithm is presented in Section \ref{Section III The proposed algorithm}.
Experiments are presented in Section \ref{Section IV Experiments}.
In Section \ref{Section V Relate Work}, local community detection methods, multiple community detection methods, and multiple local community methods are briefly introduced.  Finally, the paper is summarized in Section \ref{conlusion}.

\section{Problem Statement} \label{Section II Problem Statement}
\textbf{Data model.} Multiple networks consist of several networks, denoted by $G=\{G_1, G_2, ..., G_m, S\}$, where $G_w=(V^w, E^w)$ represents the network $w$ and $m$ represents the number of networks. Here, $V^w$ denotes the set of nodes in network $w$ and $E^w \in \mathbb{R}^{|V^w|*|V^w|}$ represents the adjacency matrix of network $w$. $S = \{S^{w \to w'}|w,w' \in [1,m]\}$ denotes the inter-layer matrix set, where $S^{w \to w'}$ denotes the inter-layer matrix between the networks $w$ and  $w'$.

Multiple networks often categorize into multi-view networks and multi-domain networks \cite{cite6_10.1145/3394486.3403069}. Examples of these two types of networks are shown in Figs. \ref{multi-view networks} and \ref{multi-domain networks}, respectively. 
If nodes in different networks from the same node set can be either identical or partially overlapping and edges in different networks represent different relationships, then this type of network is regarded as multi-view network.
Correspondingly, $S^{w \to w'}$ is the identity matrix. 
If different networks have different sets of nodes from different domains, then this type of network is regarded as multi-domain network.
There exist complex inter-layer edges between different networks, such that $S^{w \to w'} \in \mathbb{R}^{|V^w|*|V^{w'}|}$ denotes the inter-layer matrix between the network $w$ and network $w'$, where $S^{w \to w'}_{ij} = 0$ indicates there is no inter-layer edges between node \textit{i} in network $w$ and node \textit{j} in network $w'$.

\noindent  
\textbf{Problem Statement.} (Local Community Detection in Multiple Networks, LCDMN). Given the seed node $s$ and the multiple networks $G=\{G_1, G_2, ... , G_m, S\}$. The task aims to leverage information from multiple networks to identify the community containing the seed node in each network while ensuring that communities across different networks are as consistent as possible.

\noindent
\begin{table}[!t]
    \centering
    \caption{Important notations.}
    \label{notations}
    \begin{tabular}{p{0.16\columnwidth}<{\centering}  p{0.71\columnwidth}}
    \toprule
         Symbol & Definition and description \\ 
    \midrule
         $m$ & number of networks\\
         $G_w = (V^w,E^w)$ &  network $w$  with nodes set $V^w$ and adjacency matrix $E^w$\\
         $V^w_v$  &   visited nodes set  in network $w$  \\
         $z^w$ &  local affiliation vector in network $w$ \\
         $\delta_w$ &  weight of network $w$ \\
         $U$ &  unified affiliation of multiple networks\\
         $S^{w \to w'}$ &  inter-layer matrix between networks $w$ and $w'$\\
         $P^{n \to w}$ &  local inter-layer matrix between networks $n$ and $w$\\
         $||X||_F$ &  frobenius norm of X\\
         $\beta$ &  regularization parameter\\
         $t$ &  number of nodes map from the seed node\\
    \bottomrule 
    \end{tabular}
\end{table}

\section{The proposed algorithm} \label{Section III The proposed algorithm}
\begin{figure*}
    \centering
    \includegraphics[width=1\textwidth]{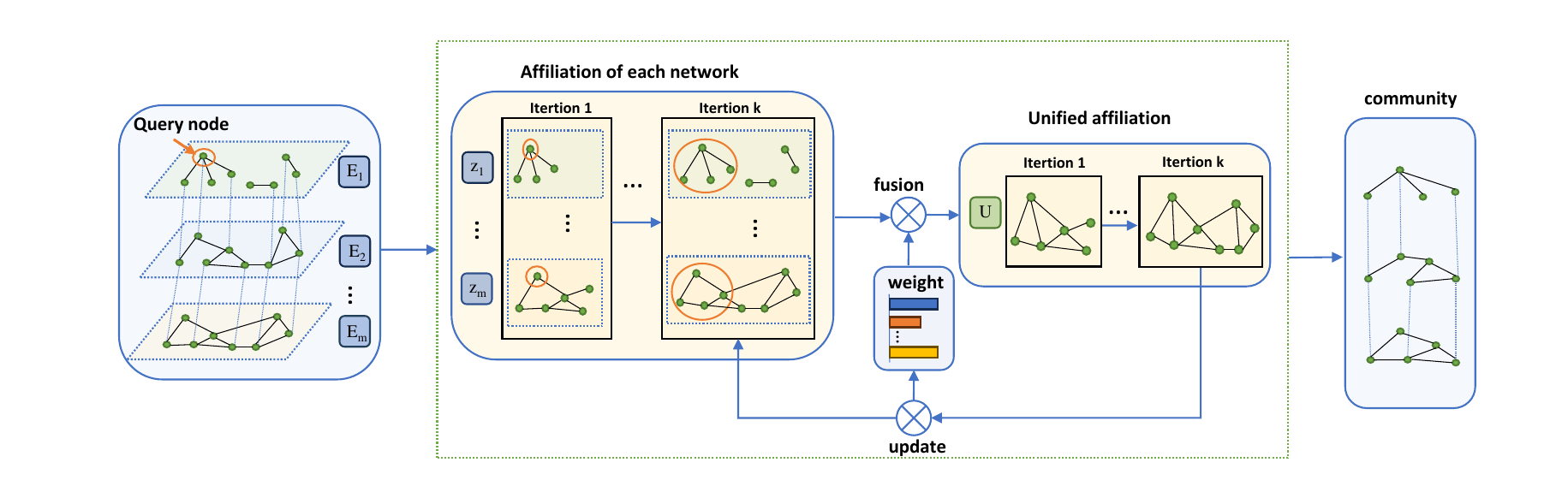}
    \caption{The diagram of LAMA}
    \label{The framework of our proposed LAMA}
\end{figure*}

In this section,  to mining local communities, we introduce a Local Algorithm for Multiple networks with node Affiliation, termed LAMA.
LAMA aims to obtain node affiliations by maximizing the quality of communities within each network while ensuring consistency in community structures across multiple networks.
Important notations are summarized in Table \ref{notations}.

Fig. \ref{The framework of our proposed LAMA} illustrates the diagram of LAMA.
LAMA consists of two steps: optimizing node affiliations and expanding local communities based on node affiliations. 
Specifically, the node affiliations is obtained by maximizing the community quality of each network and the consistency of the node affiliations of multiple networks. After obtaining the node affiliations, LAMA adds the node with high affiliation to the community to expand the community.

\subsection{Node Affiliation}

\subsubsection{Affiliation Learning of Each Network} 
The visited node $i$ in network $w$ has an affiliation $z_i^w$ and 
$z^w$ is composed of the affiliation of each visited node in the network $w$.
The node's affiliation indicates the possibility of its belonging to the community $C$. 
The nodes within the community exhibit high affiliation and the nodes outside the community show low affiliation.

The high quality of the community in each network indicates that the inner nodes of the community are closely connected and the outer nodes are sparsely connected with the nodes in the community.  
Here, we use the difference in affiliation between nodes to measure the sparsity of connections between nodes inside and outside the community, namely 
$\sum\limits_{i,j \in V^w_v} e_{ij}^w|z_i^w-z_j^w|$
where $V^w_v$ denotes the visited node set of network $w$ and $e_{ij}^w \in $ $E^w$. It correlates to the community's external edge count. 
Correspondingly, we use the closeness of affiliations between nodes to gauge the tightness among nodes within a community, namely $\sum\limits_{i,j \in V^w_v} e_{ij}^w [1-|z_i^w-z_j^w|]$. It correlates to the community's internal edge count.
Due to a small difference in the affiliations of two nodes outside the community, $\sum\limits_{i,j \in V^w_v} e_{ij}^w [1-|z_i^w-z_j^w|]$ incorrectly treats the edge between these two nodes as an internal edge.
To eliminate the influence of edges between nodes outside the community on  $\sum\limits_{i,j \in V^w_v} e_{ij}^w [1-|z_i^w-z_j^w|]$, an additional penalty term $\sum\limits_{i,j \in V^w_v} e_{ij}^w [max(z_i^w, z_j^w)-1]$ is adopted.
Based on the above analysis, the quality of the community with affiliation in network $w$  is calculated as:
\begin{equation} 
    \frac{\sum\limits_{i,j \in V^w_v} e_{ij}^w [1-|z_i^w-z_j^w|]+[max(z_i^w, z_j^w)-1]} {\sum\limits_{i,j \in V^w_v} e_{ij}^w|z_i^w-z_j^w|}.\label{local modularity for single network}
\end{equation}
A smaller value indicates better community quality. Based on Eq. (\ref{local modularity for single network}), the quality of the communities of multiple networks is calculated as:
\begin{equation}
    \sum_{w=1}^m \frac{\sum\limits_{i,j \in V^w_v} e_{ij}^w [1-|z_i^w-z_j^w|]+[max(z_i^w, z_j^w)-1]} {\sum\limits_{i,j \in V^w_v} e_{ij}^w|z_i^w-z_j^w|}, \label{local modularity for multiple network}
\end{equation}
where \textit{m} denotes the number of networks.
To enhance the quality of communities, it is essential to maximize Eq. (\ref{local modularity for multiple network}). 
To facilitate the solution, we convert the maximization problem described by Eq.  (\ref{local modularity for multiple network}) into a minimization problem by taking the reciprocal of Eq.  (\ref{local modularity for multiple network}).
Additionally, to reduce the risk of overfitting, a regularization term is incorporated. 
Consequently, Eq. (\ref{local modularity for multiple network}) is reformulated as:

\begin{equation}
\begin{split}
    & min \sum_{w=1}^m \frac{\sum\limits_{i,j \in V^w_v} e_{ij}^w|z_i^w-z_j^w|}{\sum\limits_{i,j \in V^w_v} e_{ij}^w \{[1-|z_i^w-z_j^w|]+[max(z_i^w, z_j^w)-1]\}} \\
    & + \beta \sum_{v=1}^m \|z^w\|_F^2,  \label{affiliation learning}
\end{split}
\end{equation}
where $\beta$ represents the regularization parameter.

The higher the number of internal edges of the community (i.e., the larger the denominator) and the lower the number of external edges (i.e., the smaller the numerator), the better the quality of the community (the smaller the value of the whole). 

For better understanding, let's take the affiliations of nodes $i$ and $j$ in network $w$, i.e., $z_i^w$ and $z_j^w$, as examples and assume $e_{ij}=1$ to explain the rationality of Eq. (\ref{affiliation learning}).

\begin{itemize}
  \item  If the difference between $z_i^w$ and $z_j^w$, i.e., $e_{ij}^w |z_i^w - z_j^w|$, is large, it indicates that one of the nodes $i$ and $j$ is likely within the community, while the other is likely outside the community. Therefore, the edge $(i, j)$ is likely an external edge of the community.
The denominator, i.e., $e_{ij}^w \{[1-|z_i^w-z_j^w|] + [max(z_i^w, z_j^w)-1]\}$,  is small implies that the edge is less likely to be an internal edge of the community. 
For example, assume $z_i^w$ = 0.9 and $z_j^w$ = 0.2.
$e_{ij}^w|z_i^w-z_j^w|$ = 0.7, $e_{ij}^w \{[1-|z_i^w-z_j^w|]+[max(z_i^w, z_j^w)-1]\}$ = 0.3 - 0.1 = 0.2. 
The edge between node $i$ and node $j$ contributes 0.7 (0.2) to the numerator (denominator), representing the possibility of the external (internal) edge of the community.
  \item If the difference between $z_i^w$ and $z_j^w$, i.e., $e_{ij}^w |z_i^w - z_j^w|$, is small and both are close to 1 (or 0), it indicates that both nodes $i$ and $j$ is likely within the community (outside the community). Therefore, the edge $(i, j)$ is less likely an external edge of the community.
The denominator, i.e., $e_{ij}^w \{[1-|z_i^w-z_j^w|] + [max(z_i^w, z_j^w)-1]\}$,  is large (small) implies that the is less likely to be an internal edge of the community (external edge of the community). 
For example, assume $z_i^w$ = 0.9 and $z_j^w$ = 0.8.
$e_{ij}^w|z_i^w-z_j^w|$ = 0.1, $e_{ij}^w \{[1-|z_i^w-z_j^w|] + [max(z_i^w, z_j^w)-1]\}$ = 0.9 - 0.1 = 0.8. 
The edge between node $i$ and node $j$ contributes 0.1 (0.8) to the numerator (denominator), representing the possibility of the external (internal) edge of the community.
Here, the case that $z_i^w$ and $z_j^w$ are both close to 0 is omitted.
\end{itemize}
\subsubsection{Unified Affiliation Learning}


In multiple networks, nodes in network $w$ are connected to nodes in network $w'$, indicating that there is consistency between the community structure in network $w$ and that in network $w'$.
The more edges between the networks $w$ and $w'$, the higher the consistency of the communities.
This consistency implies that the difference between $z^w$ and $z^{w'}$ is small, calculated as:
\begin{equation}
      \| z^w - S^{w \to w'} * z^{w'} \|_F^2 , \label{eql:unified}
\end{equation}
where $S^{w \to w'}$ represents the inter-layer edges between network $w$ and network $w'$. %

To maintain consistency between communities between networks, the intuitive idea involves limiting the differences in the $z$ values between any two networks; however, this method incurs considerable complexity.
To simplify calculations, we construct a unified affiliation, denoted as $U$, and achieve consistency by limiting the proximity of $U$ and $z^{w'}$ of the network $w'$. 
To calculate the degree of consistency between $U$ and $z^{w'}$, it is assumed that $U$ and the network  containing the seed node are from the same perspective, namely network $n$.
That is, the consistency of $U$ and $z^{w'}$ is calculated by calculating $\| U - S^{n \to w'} * z^{w'} \|_F^2$.
Furthermore, since the consistency between $z$ of the network and $U$ varies,  different weights are assigned to each network. Through the above analysis, the difference between unified affiliation and $z$ of  networks can be derived:

\begin{equation}
     \sum_{w=1}^m \delta_w \| U - S^{n \to w} * z^w \|_F^2 , \label{non-local unified affiliation learning}
\end{equation}
where $\delta_w$ 
denotes the weight of network $w$. 

For each network $w$, only nodes around seed nodes are visited, denoted as $V^{w}_v$. For the visited nodes $V^{w}_v$  and the visited nodes $V^{w'}_v$, it is not necessary to utilize  $S^{w \to w'}$ between network $w$ and network $w'$.
To avoid accessing unnecessary information in $S^{w \to w'}$, the rows and columns associated with nodes related to the visited nodes are extracted from $S^{w \to w'}$ to form a new matrix $S_l^{w \to w'}$.
The nodes related to the visited nodes refer to the visited nodes of the network $w$ and the nodes that map the visited nodes of other networks to the network $w$. This is to avoid the nodes in $S_l^{w \to w'}$ not covering the visited nodes.
In addition, the node in network $n$ may be connected to the node in network $w$ through nodes in other networks.
To make full use of inter-layer information, we obtain all local inter-layer edges between networks $n$ and  $w$, denotes $\hat{S^{n \to w}}$:
\begin{equation}
\begin{split}
    & \hat{S^{n \to w}}= \{ \prod_{i=0}^{k} S_l^{a_i \to a_{i+1}} | (a_0, a_1, \dots, a_{k+1}) \in T_{n \to w}, \\
    & a_0 = n, a_{k+1} = w \},
    \label{eql:s}
\end{split}
\end{equation}
where $T_{n \to w}$ denotes the set of all possible paths from network $n$ to network $w$ and each path $(a_0, a_1, \dots, a_{k+1})$ is represented as a sequence of networks. 
For example, $(n, a, w)$ denotes the path from network $n$ to network $w$ via an intermediary network $a$. This path facilitates the derivation of $ S_l^{n \to a}*S_l^{a \to w}$, which represents the connections from network $n$ to network $w$ through intermediary network $a$.

Through Eq. (\ref{eql:s}), we can obtain the inter-layer matrix of networks $n$ and $w$ through different paths. Based on these inter-layer matrices, we define the local strongest inter-layer matrix $P^{n \to w}$ of network $n$ and $w$. The element $P^{n \to w}_{ij}$ in $P^{n \to w}$ is calculated as: 
\begin{equation}
    P^{n \to w}_{ij} = \max \left\{ S'_{ij} | {S'} \in \hat{S^{n \to w}} \right\}. \label{localP}
\end{equation}
In particular, $P^{n \to w}$ is the identity matrix when $w = n$. 
For the multi-view network, $P^{n \to w}(1 \le w \le m)$ is the identity matrix.
Next, we replace $S^{n \to w}$ in Eq. (\ref{non-local unified affiliation learning}) with the obtained $P^{n \to w}$, and Eq. (\ref{non-local unified affiliation learning}) is rewritten as:
\begin{equation}
    min \sum_{w=1}^m \delta_w \| U - P^{n \to w} * z^w \|_F^2 . \label{unified affiliation learning}
\end{equation}

\subsubsection{Optimal Affiliation Learning}
By joining Eq. (\ref{affiliation learning}) and Eq. (\ref{unified affiliation learning}) to obtain  Eq. (\ref{optimal affiliation learning}), we can co-learn the affiliation $z^w$ of each network, the unified affiliation $U$, and the weights $\delta_w$ of each network: 
\begin{equation}
\begin{split}
    & min \sum_{w=1}^m \frac{\sum\limits_{i,j \in V^w_v} e_{ij}^w|z_i^w-z_j^w|}{\sum\limits_{i,j \in V^w_v} e_{ij}^w [1-|z_i^w-z_j^w|]+[max(z_i^w, z_j^w)-1]} \\
    & + \beta \sum_{w=1}^m \|z^w\|_F^2 + \sum_{w=1}^m \delta_w \| U - P^{n \to w} * z^w \|_F^2.  \label{optimal affiliation learning}
\end{split}
\end{equation}

\subsection{Optimization Strategy}
The optimization of the variables $\{z^w\}_{w=1}^m$, $\{\delta_w\}_{w=1}^m$, and $U$, as well as the expansion of the local community, are performed iteratively.
The optimization of $\{z^w\}_{w=1}^m$, $\{\delta_w\}_{w=1}^m$, $U$ adopts an alternating iteration strategy, i.e., when the other two variables are fixed, the remaining one is updated.

\subsubsection{Fix \texorpdfstring{$\{\delta_w\}_{w=1}^m$ and $U$, Update $\{z^w\}_{w=1}^m$}{fix delta w from 1 to m, U, update z w from 1 to m}}
When we fix $\{\delta_w\}_{w=1}^m$ and $U$, it's easy to find that each network's affiliation $z^w$ is updated independently, but not coupled together. 
So, we can update $z^w$ for each network one by one, formulated as:
\begin{equation}
\begin{aligned}
    & min \frac{\sum\limits_{i,j \in V^w_v} e_{ij}^w|z_i^w-z_j^w|}{\sum\limits_{i,j \in V^w_v} e_{ij}^w [1-|z_i^w-z_j^w|]+[max(z_i^w, z_j^w)-1]} \\
    & + \beta  \|z^w\|_F^2 + \delta_w \| U - P^{n \to w} * z^w \|_F^2.\\  \label{update z}
\end{aligned}
\end{equation}


Given that Eq. (\ref{update z}) is non-differentiable, it is not feasible to obtain extreme values through differentiation. Moreover, there is no need to achieve an exact solution for $z_i^w$; it is sufficient for the values of $z_i^w$ to indicate whether the node belongs the community or not.
A simple approach is employed for the solution.
By uniformly sampling 21  values between 0 and 1, the value that minimizes Eq. (\ref{update z}) is selected as $z_i^w$.

\subsubsection{Fix \texorpdfstring{$\{z^w\}_{w=1}^m$ and $U$, Update $\{\delta_w\}_{w=1}^m$}{fix z from 1 to m, U, update delta w from 1 to m}}
When we fix $\{z^w\}_{w=1}^m$ and $U$ to update $\{\delta_w\}_{w=1}^m$, the Eq. (\ref{optimal affiliation learning}) can be transformed into:
\begin{equation}
    \min \sum_{w=1}^m \delta_w \| U - P^{n \to w} * z^w \|_F^2.  \label{update w}
\end{equation}
Similar to the approach in \cite{cite66_9186335}, by processing Eq. (\ref{update w}), we can obtain the following result:
\begin{equation}
    \delta_w = \frac{1}{2 * \sqrt{\| U - P^{n \to w} * z^w \|_F^2}}.  \label{value w}
\end{equation}
Proof: Define the function:

\begin{equation}
    \min_{U} \sum_{w=1}^m \sqrt{\| U - P^{n \to w} * z^w \|_F^2}.  \label{auxiliary function}
\end{equation}

Taking the derivative of Eq. (\ref{auxiliary function}) for $U$ and setting the derivative to zero, we have:
\begin{equation}
    \sum_{w=1}^m \hat{\delta_w} \frac{\partial \| U - P^{n \to w} * z^w \|_F^2 }{\partial U} = 0,  \label{setting the derivative to zero}
\end{equation}
where
\begin{equation}
    \hat{\delta_w} = \frac{1}{2 * \sqrt{\| U - P^{n \to w} * z^w \|_F^2}}. \label{lambda w value}
\end{equation}
Also, taking the derivative of Eq. (\ref{update w}), we have:
\begin{equation}
    \sum_{w=1}^m \delta_w \frac{\partial \| U - P^{n \to w} * z^w \|_F^2 }{\partial U} = 0. \label{Eq7 derivative to zero}
\end{equation}
From Eq. (\ref{setting the derivative to zero}) and Eq. (\ref{Eq7 derivative to zero}), we can get the same solution to Eq. (\ref{update w}) and Eq. (\ref{auxiliary function}) if $\hat{\delta_w} = \delta_w$. At this point, the solution of weight $\delta_w$ is shown in Eq. (\ref{value w}).
Eq. (\ref{setting the derivative to zero}) cannot be solved directly, due to $\delta_w$ depending on the $U$ and $z^w$. If $\delta_w$ is set stationary, the solution of Eq. (\ref{setting the derivative to zero}) can be regarded as the solution of Eq. (\ref{update w}). The updated $U$ through Eq. (\ref{setting the derivative to zero}) will be employed to update $\delta_w$ via Eq. (\ref{value w}). So, Eq. (\ref{auxiliary function}) can be solved through an iterative strategy, and the converged $U$ and $\delta_w$ will be optimal if the iterative optimization. Also, $\delta_w$ can turn to an optimal value via Eq. (\ref{value w}). Therefore, the Eq. (\ref{update w}) can be transformed into Eq. (\ref{auxiliary function}) and the value of $U$ and $z^w$ can be obtained through the last iteration when $\delta_w$ is determined by Eq. (\ref{value w}).

\subsubsection{Fix \texorpdfstring{$\{z^w\}_{w=1}^m$ and $\{\delta_w\}_{w=1}^m$, Update $U$}{fix z from 1 to m, delta w from 1 to m, update U}}
When we fix $\{z^w\}_{w=1}^m$ and $\{\delta_w\}_{w=1}^m$ to update $U$, the Eq. (\ref{optimal affiliation learning}) can transformed into:
\begin{equation}
    min  \sum_{w=1}^m \delta_w \| U - P^{n \to w} * z^w \|_F^2. \label{trans. optimal affiliation learning}
\end{equation}
Eq. (\ref{trans. optimal affiliation learning}) is a sum of squared norms about $U$, which is a convex function. For convex functions, every local minimum is inherently a global minimum, and any point where the gradient is zero is acknowledged as a global minimum\cite{cite71_DBLP:journals/axioms/KhanASN23}. Therefore, we solve $U$ by setting its gradient to zero:
\begin{equation}
    \sum_{w=1}^m \frac{\partial \delta_w \| U - P^{n \to w} * z^w \|_F^2 }{\partial U} = 0.  \label{partial u}
\end{equation}
From Eq. (\ref{partial u}), we have:
\begin{equation}
    u_i = \sum \delta_w * P^{n \to w} * z_i^w. \label{update ui}
\end{equation}

\subsection{Expanding Local Community} 
As the community expands, LAMA continues to visit more nodes in each network and the visited nodes are divided into core node set $C$, boundary set $N$, and shell set $NN$.
Here, $C$ contains nodes with high affiliation and a high probability of belonging to the community, 
$N$ consists of neighbor nodes of $C$,
and $NN$ comprises the neighbor nodes of $N$ and $C$.

For the network $w$, after optimizing  the variables  $\{z\}_{w=1}^m$, $\{\delta_w\}_{w=1}^m$, and $U$ in each iteration, LAMA expands  the core node set $C^w$ through  $z^w$.
Specifically, LAMA adds nodes with affiliation greater than $z_{norm}^w$ in both $C^w$ and $N^w$ to  $C^w$, where $z_{norm}^w$ is calculated as:
\begin{equation}
    z_{norm}^w = \frac{1}{|C^w| + |N^w|} \sum_{i \in  (C^w \bigcup N^w)} z_i^w.
    \label{eql:znorm}
\end{equation}
Subsequently, LAMA updates $N^w$ using the neighbor of $C^w$ and updates $NN^w$ through the neighbor nodes of $C^w$ and $N^w$.

Algorithm \ref{alg LAMA} shows the pseudo-code of LAMA. 
\begin{algorithm} [!t] 
	\caption{LAMA}     
        \label{alg LAMA}
    	\begin{algorithmic}[1] 
    	\Require network $G=\{G_1, G_2, ..., G_m\}$, seed, inter-layer matrix set $S = \{S^{w \to w'}|w, w' \in [1,m]\}$
        \Ensure  community \textit{C}   
        \State 	Initialize  $C^w$,  $N^w$ and  $NN^w$ where $w \in [1,m]$  \label{line1} 
        \State  Initialize $z^w$ for network $w$ by Eq. (\ref{affiliation learning})  where $w \in [1,m]$  \label{line2} %
        \State  Initialize  $\delta_w=1/m$ for network $w$ where $w \in [1,m]$ \label{line3}
        \State  Initialize  $\{P^w\}^m_{w=1}$  by Eq. (\ref{localP}) \label{line4}
        \State  Initialize $U$ by Eq. (\ref{update ui}) \label{line5}
         \While {C stabilized or maximum iteration reached} 
           \State Update $\{z^w\}^m_{w=1}$ by using Eq. (\ref{update z})   \label{line6}
           \State Update $\{\delta_w\}^m_{w=1}$ by using Eq. (\ref{update w})    \label{line7}
           \State Update $U$ by using Eq. (\ref{update ui})       \label{line8}
           \State Extend $C$ by using Eq. (\ref{eql:znorm})
           \State Update $N$ and $NN$      \label{line9}
           \State Adjust $\{z^w\}^m_{w=1}$, $\{P^{n \to w}\}^m_{w=1}$ and $U$ \label{line adjust}
       \EndWhile
        \State Obtain community $C^w$ for network $w$ via $U$ and  $z^w$      \label{line10}
       \State\Return community $\{C^w\}^m_{w=1}$ 
       \end{algorithmic} 
\end{algorithm}
Initially, the community $C^w$ for each network is initialized (line \ref{line1}).
For the network containing the seed node, the community is initialized as a set containing only the seed node.
For the multi-view networks, $C^w$ in network $w$ is initialized with only the seed nodes.
For the multiple networks, $C^w$ in network $w$ is initialized with the top $t$ nodes that exhibit the strongest inter-layer matrix with the seed node in $P$.
Based on the initialized $C^w$, $N^w$ and $NN^w$ are initialized for each network.
Then, LAMA initializes nodes affiliation in $C$, $N$, and $NN$ to 1, 0.5, and 0, respectively. It further adjusts  $z$ of nodes in $N$ by minimizing Eq. (\ref{affiliation learning})(line \ref{line2}). LAMA initializes the local strongest inter-layer matrix set $P$ by Eq. (\ref{localP})(line \ref{line4}) and initializes $U$ by Eq. (\ref{update ui})(line \ref{line5}).
Next, updating the variables and extending the local community are iteratively performed until $C^w$ for each network achieves stability or the maximum number of iterations is reached (lines \ref{line6}-\ref{line9}). 
Besides, LAMA adjusts the dimensions of variables $\{z^w\}^m_{w=1}$, $\{P^{n \to w}\}^m_{w=1}$ and $U$ to accommodate the nodes added during the community expansion(line \ref{line adjust}).
Finally, $\delta_w*z^w+(1-\delta_w)*U$ is computed and the nodes within $\delta_w*z^w+(1-\delta_w)*U$ exceeding the mean unified affiliation are identified as members of the local communities associated with the seed nodes for each network (line \ref{line10}).

\section{Experiments}\label{Section IV Experiments}
In this section, we first introduce the experimental setup, including the dataset, comparison algorithm, and evaluation metrics. Then the experimental results and analysis of the results are presented. 
All experiments were run on a Window 10 desktop computer (CPU: Inter(R) Core(TM) i9-12900k@ 3.19GHz, memory 64GB). LAMA is implemented in Matlab 2021a development environment.
\subsection{Experiment Setting}
\subsubsection{Dataset}
Two synthetic datasets and five real datasets are used to evaluate the performance of algorithms. 
Table \ref{datas} summarizes the statistics information of these datasets. The two synthetic datasets are PEP and PNP \cite{cite28_magnani2021community}. 
PEP and PNP are multi-view networks.
Each network encompasses three views, with each view containing one hundred nodes. The PEP dataset features communities of the same sizes, whereas the PNP dataset exhibits variability in community sizes.
The five real datasets include 3-sources \cite{cite16_liu2013multi}, BBC \cite{cite30_greene2005producing}, WebKb \cite{cite35_craven1998learning}, 6NG \cite{cite49_ni2018co}, and 9NG \cite{cite49_ni2018co}. The first three datasets are multi-view networks with the same set of nodes, while the last two are multi-domain network datasets. 
The 3-sources dataset comprises a total of 948 news stories collected from BBC, Reuters, and The Guardian between February and April 2009. It covers 416 distinct news stories, including 169 that are reported across all three sources. These are categorized under the labels ``Business", ``Entertainment", ``Health", ``Politics", ``Sports", and ``Science and Technology".
The BBC dataset leverages the BBC news corpus to create four views of documents, based on relevant text segments, containing a total of 2225 documents tagged under five subjects. 
The WebKb dataset comprises 187 documents, tagged under five categories: ``Student", ``Project", ``Course", ``Staff", and ``Faculty". It features content and citations, forming the two respective views. The 6NG and 9NG datasets are based on a 20-Newsgroup dataset constructed on two multi-domain web datasets, where nodes denote news documents and edges denote their semantic similarity.

For multi-view networks, each node is used as a seed node to obtain the community structure of that node; for multi-domain networks, each node in the first network is used as a seed node to obtain its community.

\begin{table}[!t]
  \centering
  \caption{Statistics on real-world datasets}
    \begin{tabular}{cccccc}
    \toprule
        Dataset & Node & Inside edge & Cross edge & View & Cmty \\ \midrule
        PEP & 100 & 1636 & {-} & 3 & 10 \\ 
        PNP & 100 & 2786 & {-} & 3 & 10 \\ 
        3-sources & 169 & 6312 & {-} & 3 & 6 \\ 
        BBC & 685 & 1528 & {-} & 4 & 5 \\ 
        WebKb & 187 & 1636 & {-} & 2 & 5 \\ 
        6NG & 4500 & 9000 & 20984 & 5 & 6 \\ 
        9NG & 6750 & 13500 & 31480 & 5 & 6 \\ 
    \bottomrule
    \end{tabular}%
  \label{datas}
\end{table}%

\subsubsection{Comparison Algorithms}

To validate the performance of  LAMA, we compare it with three local community detection methods on multi-view networks (i.e., ML-LCD, PLCDM, and RWM) and one local community detection method on multi-domain networks  (i.e., RWM).

1) ML-LCD\cite{cite50_interdonato2017local}: ML-LCD aims to optimize an objective function, which measures the link density inside and outside a local community. It employs a greedy strategy to efficiently identify local communities, starting from a seed node on multiple networks.

2) PLCDM\cite{cite27_pournoor2021propagation}: 
PLCDM initiates a random walker to obtain the community to which the seed node belongs.
It stochastically transitions to neighboring nodes within the same network or to nodes in other networks, taking into account both intra-layer and inter-layer transition probabilities.

3) RWM\cite{cite6_10.1145/3394486.3403069}: 
RWM launches a random walker for each layer of the network, which is confined to random walkers within its respective layer. The node transition probabilities are influenced by both the current network and the transition probabilities from other layers.

\subsubsection{Metrics}
Three evaluation metrics recall, precision, and fscore\cite{cite70_10414283} are used in this experiment to measure the results of the detected communities. 
To evaluate the performance of algorithms on multiple networks with ground truth, we use recall, precision, and fscore.
For convenience, let $T$ represent the ground truth community containing the seed node, and $C$ represent the community found by the method with the seed node as the starting node.
Based on $C$ and $T$, recall, precision, and fscore in a single network are described as follows:

\begin{equation}
    recall = \frac{|C \cap T|}{|T|}, \label{recall}
\end{equation}

\begin{equation}
    precision = \frac{|C \cap T|}{|C|},  \label{precision}
\end{equation}

\begin{equation}
    fscore = 2 * \frac{precision*recall}{precision+recall} . \label{f1score}
\end{equation}


For each seed node, the algorithm detects a local community within each network. The indicator of the detected community in each network is calculated, and the average of the indicator values of all networks is taken as the indicator value of the seed node. 
Then, each node in the network is used as a seed node and the average of the indicator values of all seed nodes is calculated as the indicator of the algorithm.  recall, precision, and fscore
Finally, we calculate the indicators recall, precision, and fscore as described above.


\subsection{Results}
The results on multi-view networks and multi-domain networks datasets are shown in Tables \ref{result of multiview network} and \ref{result of multi domain network}, respectively.
\subsubsection{Results on Multi-view Networks}
Table \ref{result of multiview network} shows the results on the multi-view networks. 
In terms of fscore, LAMA achieves the best performance on most datasets. For the BBC dataset, the performance of LAMA is second only to RWM  and significantly better than the other methods.
The relatively low recall value of PLCDM is due to the community detected by PLCDM being relatively small. 
RWM and LAMA perform well on multi-view networks, which indicates that the complementary information from other networks can effectively help to improve the quality of communities.
Compared with RWM, LAMA additionally makes the communities between multiple views as consistent as possible by unifying the affiliations, which makes LAMA perform better than  RWM.
LAMA outperforms ML-LCD on all datasets. 
The reason is that the layers of weights in ML-LCD are designated by either an unsupervised or supervised approach, and the detected communities rely on these specified layer weights, which may not accurately reflect the network's actual conditions.
In contrast,  LAMA adaptively learns the weights of each view during the expansion process.


\subsubsection{Results on Multi-domain Networks}
Table \ref{result of multi domain network} shows the performance of RWM and LAMA on multi-domain networks.
In terms of fscore and recall, LAMA achieves better performance than RWM on all datasets. 
The reason may be as follows.
The decay factor $\lambda$ needs to be set in RWM and affects the similarity of local structures between networks.
The default  $\lambda$ value may not suit the actual conditions of the 6NG and 9NG datasets, which makes the community size obtained by RWM smaller than the ground truth community, resulting in low recall.
LAMA constrains the consistency between communities across multiple networks through unified affiliations and the weights of individual networks are adaptively determined, which makes LAMA superior to RWM in terms of performance.  


\begin{table}[!t]
  \centering
   \caption{Results on multi-view networks. }
    \begin{tabular}{cccccc}
   \toprule
Datasets & Metrics           & PLCDM     & MLLCD & RWM & LAMA \\
    \midrule
    \multirow{3}{*}{PEP} & recall & 0.8860  & 0.8920 & 0.8970 & \textbf{1}  \\
          & precision & 0.9567  & 0.8900  & 0.8930 & \textbf{0.9964} \\
         & fscore & 0.9167  & 0.8910 & 0.8618 & \textbf{0.9981}  \\
    \midrule
    \multirow{3}{*}{PNP} & recall & 0.8027  & 0.8903 & 0.9597 & \textbf{0.9910}  \\
          & precision & \textbf{0.8931}  & 0.8638  & 0.6806 & 0.8457 \\
         & fscore & 0.8224  & 0.8698 & 0.7031 & \textbf{0.8788}  \\
    \midrule
    \multirow{3}{*}{3sources} & recall & 0.4217  & 0.5881 & \textbf{0.9562} & 0.8548  \\
          & precision & \textbf{0.7782}  & 0.7318  & 0.5131 & 0.6841 \\
         & fscore & 0.5104  & 0.6187 & 0.6044 & \textbf{0.7152}  \\
    \midrule
    \multirow{3}{*}{webkb} & recall & 0.1411  & 0.2554 & \textbf{0.4452} & 0.4124  \\
          & precision & 0.4916  & 0.4982  & 0.4177 & \textbf{0.5193} \\
         & fscore & 0.1980  & 0.2855 & 0.3774 & \textbf{0.3889}  \\
    \midrule
    \multirow{3}{*}{BBC} & recall & 0.2206  & 0.2214 & \textbf{0.6648} & 0.5630  \\
          & precision & 0.7667  & 0.7960  & 0.6724 & \textbf{0.9261} \\
         & fscore & 0.3296  & 0.3204 & \textbf{0.6414} & 0.6053  \\

    \bottomrule
    \end{tabular}%
    \label{result of multiview network}
\end{table}%

\begin{table}[!t]
  \centering
   \caption{Results on multi-domain networks. }
    \begin{tabular}{cccc}
   \toprule
Datasets & Metrics            & RWM & LAMA \\
    \midrule
    \multirow{3}{*}{6NG} & recall & 0.6188  & \textbf{0.6617 }  \\
          & precision & 0.5850 & \textbf{0.5956}   \\
         & fscore & 0.5325  & \textbf{0.6158}   \\
    \midrule
    \multirow{3}{*}{9NG} & recall & 0.2528  & \textbf{0.5591}  \\
          & precision & \textbf{0.4509}  & 0.3348  \\
         & fscore & 0.2261 & \textbf{0.4047}   \\
    \bottomrule
    \end{tabular}%
    \label{result of multi domain network}
\end{table}%

\subsection{Discussion} 
We conduct experiments on the 6NG dataset to explore the effects of parameters $t$ and the regularization factor $\beta$ on the performance of LAMA.
One hundred nodes in the 6NG dataset are selected as seed nodes to extend the local community respectively and indicator fscore is adopted to measure the results.

\subsubsection{Effect of Parameter \texorpdfstring{$t$}{t}}
For multi-domain networks, parameter $t$ determines the number of nodes in the initial community that map from the seed node to other networks.
For multi-view networks, each network contains exactly one node that corresponds to the seed node, thereby obviating the need for $t$.
Here, $t$ takes the value $t=2*k+1, k \in [0,9]$. 
The experimental results for various values of $t$ on the 6NG dataset are depicted in Fig. \ref{The study of Parameter t on 6-NG}\subref{The study of parameter t}.

Fig. \ref{The study of Parameter t on 6-NG}\subref{The study of parameter t} demonstrates that fscore exhibits an upward trend as $t$ increases, reaching its peak when $t$ equals 11, after which it demonstrates a downward trend with further increases in $t$.
This is because the seed node in network $w$ may be connected to nodes in several communities in network $w'$ via inter-layer edges.
The parameter $t$ influences the proportion of nodes in the initial community from different communities, thereby impacting the detected communities.
Take Fig. \ref{The study of Parameter t on 6-NG}\subref{A example of parameter t study}  as an example to explain the impact of different $t$ values on the results.
Community $C_1$ in network $w$ corresponds to community $C_2$ in network $w$.  
However, seed node $A_1$ in community $C_1$ in network $w$ has inter-layer edges with nodes $P_1$ (weight 0.6), $P_2$ (weight 0.3), $P_4$ (weight 0.4) in community $C_2$ as well as $P_7$ (weight 0.6) and $P_9$ (weight 0.1) in community $C_3$ in network $w'$.
Under different $t$ values, nodes in the initial community  are as follows:


\begin{figure}[!t]
\centering
\subfloat[Fscore of Different $t$]
{
\includegraphics[width=0.9\columnwidth]{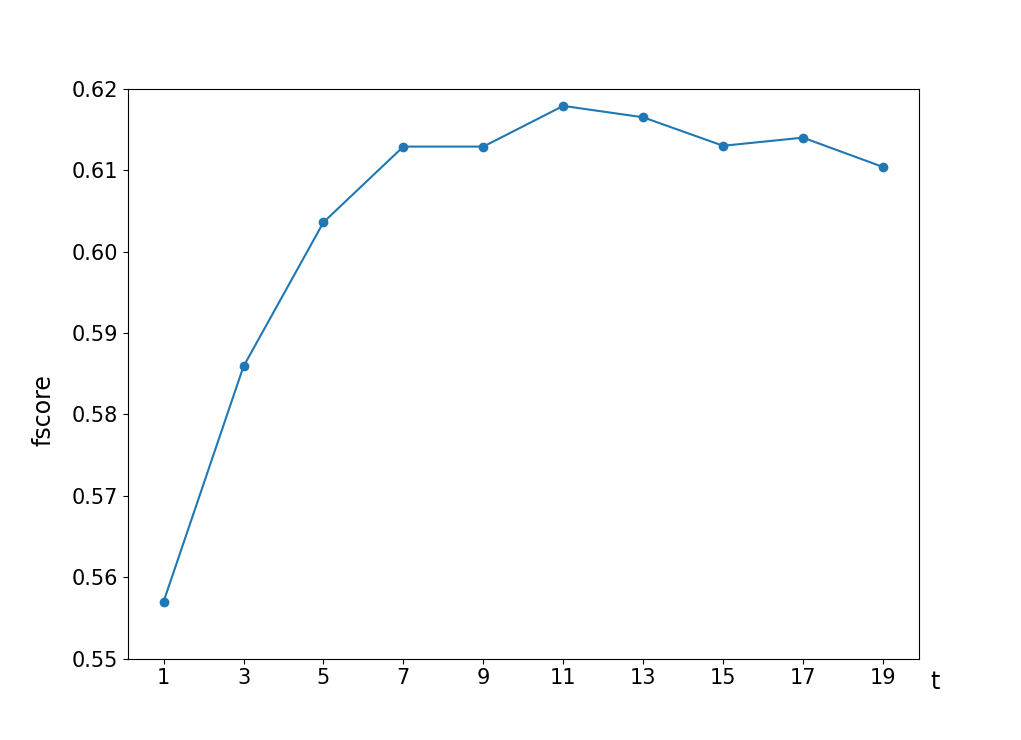}\label{The study of parameter t}}

\subfloat[Initial communities for different $t$]
{
\includegraphics[width=0.9\columnwidth]{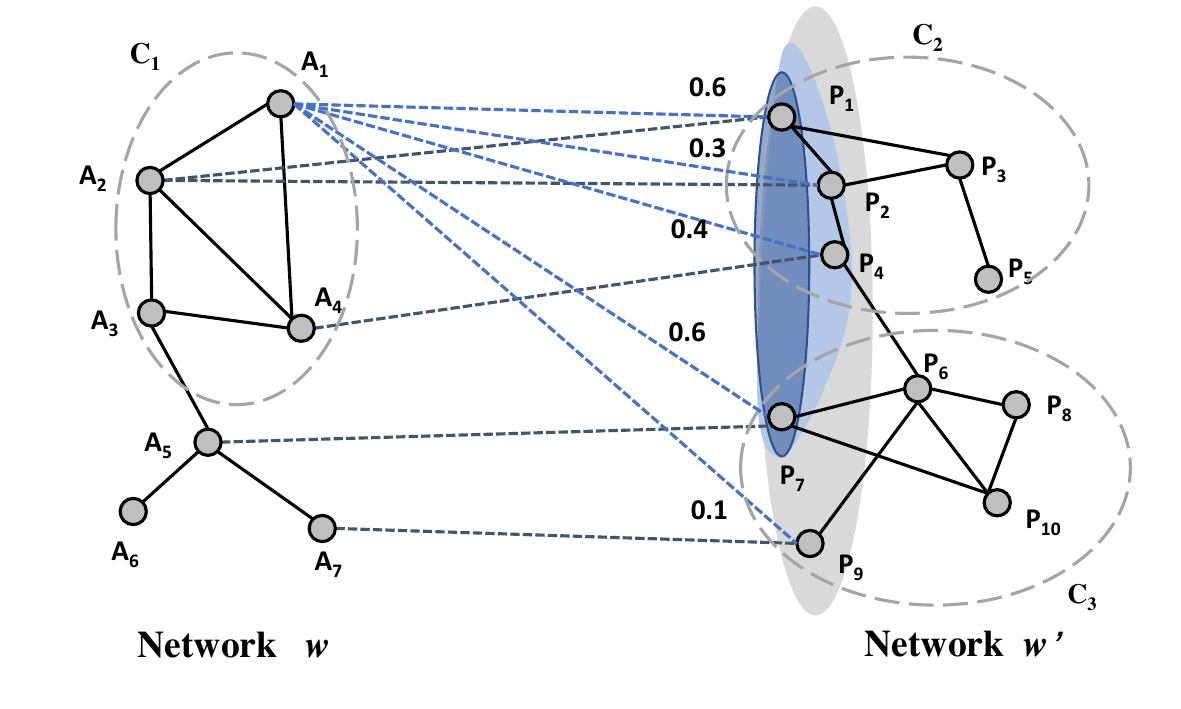}\label{A example of parameter t study}}
\caption{Effect of Parameter $t$ on 6-NG}
\label{The study of Parameter t on 6-NG}
\end{figure}

\begin{figure}
    \centering
    \includegraphics[width=0.9\columnwidth]{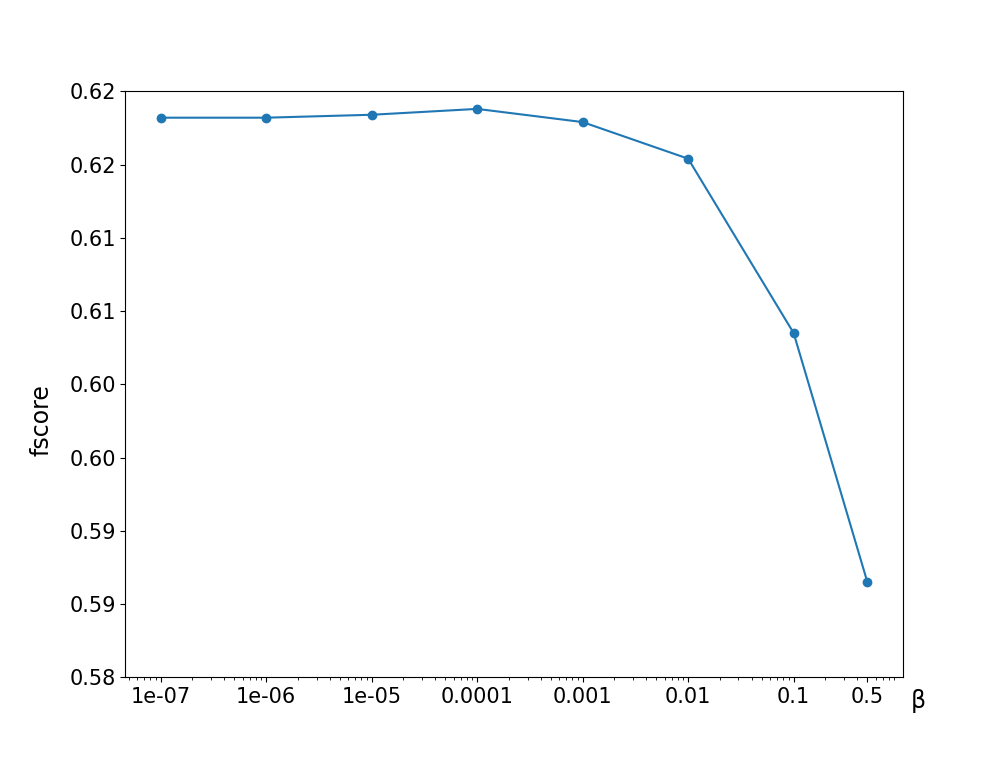}
    \caption{Effect of Parameter $\beta$ on 6-NG}
    \label{The study of Parameter beta on 6-NG}
\end{figure}

\begin{itemize}

 \item  When $t$ is small, such as $t=2$, the nodes mapping from $A_1$ are $P_1$ and $P_7$, which belong to communities $C_2$ and $C_3$, respectively.
Using $P_1$ and $P_7$ as seed nodes to expand the two communities simultaneously results in inaccurate community detection.


 \item  When $t$ increases from a small value to an optimal value, i.e., $t= 4$,  nodes that map from $A_1$ include $P_1$, $P_2$, $P_4$,  and $P_7$.  
As $t$ increases, the proportion of nodes from $C_2$ within the initial community is enhanced, while the proportion of nodes such as $P_7$ outside $C_2$  is diminished. This reduces the impact of inter-layer noise and improves the performance.



 \item  When $t$ reaches the optimum and continues to increase,  nodes like $P_2$ outside community $C_2$ are added to the initial community, diminishing the detected community's quality. Since nodes in $C_2$ predominate the initial community, the overall community quality remains relatively stable, causing performance to fluctuate.

\end{itemize}



\subsubsection{Effect of Parameter \texorpdfstring{$\beta$}{beta}}
The parameter $\beta$ is a regularization term parameter in Eq. (\ref{affiliation learning}). Here, $\beta$ takes the value $\beta = 10^{-7},10^{-6},...,10^{-1},0.5$. The experimental results for various values of $\beta$ on the 6NG dataset are depicted in Fig. \ref{The study of Parameter beta on 6-NG}

Fig. \ref{The study of Parameter beta on 6-NG} demonstrates that the fscore increases with smaller values of $\beta$, peaking at $\beta = 10^{-4}$. Beyond this value, the fscore begins to decline. This suggests that the regularization term parameter $\beta$ plays a beneficial role in community expansion and enhances generalization capabilities. However, too large values of $\beta$ disproportionately influence the initialization and updating of affiliations, as outlined in Eqs. (\ref{affiliation learning}) and (\ref{update z}). Such dominance adversely impacts the evaluation of community quality, thereby leading to a decline in performance.



\subsection{Case Study}

DBLP is a representative dataset for the multi-domain network (with flexible nodes and edges). We conduct a case study on DBLP \cite{cite69_10.1145/1401890.1402008} to show the detected local communities by LAMA. DBLP contains a co-author network and a citation network.  In the co-author network, nodes represent authors and edges denote collaborative relationships between them. In the citation network, nodes represent papers, with edges indicating citations from one paper to another.
Since no ground truth about the relevant local communities in these two networks.
So, we selected "Jie Tang" as the seed node, a researcher whose main research interests include data mining, the quality of the community is judged by observing whether the detected community nodes are related to Tang Jie's research direction. DBLP was collected in July 2016 when he was working at Tsinghua University.


Table \ref{case study} shows the author community from the co-author network and the paper community from the citation network. Due to space limitations, we do not show detailed information in the author community as well as that in the paper community, we list part authors and list venues where these papers were published.
For example, "KDD(17)" indicates that 17 papers are published on KDD in the paper community. 
Specifically, the detected authors are mainly from Tang Jie's group(Zhichun Wang, Juanzi Li, Jamal Yousaf et al.) and those who collaborated with Tang Jie's group(Xiaowen Dai, Martin A. Ferman et al.).
The papers in the paper communities are mainly published in conferences that related to data mining.
The results indicate that our method successfully identifies local communities with practical significance from DBLP.

\begin{table}[!t]
  \centering
   \caption{Results on DBLP dataset. }
    \begin{tabular}{cc}
   \toprule
    Author community  & Paper community \\
    Jie Tang & KDD(17) \\
    Zhichun Wang & CoRR(16) \\
    Juanzi Li & CIKM(16) \\
    Xiaowen Dai  &  ICDM(12) \\
    Jamal Yousaf & WWW(9)  \\
    Martin A. Ferman & AAAI(8) \\
    ...  &  ... \\
    \midrule
    \end{tabular}%
    \label{case study}
\end{table}%

\section{Relate Work} \label{Section V Relate Work}

\subsection{Local Community Detection}
Local community detection has received much attention for its capacity to rapidly identify communities containing the seed node \cite{cite2_5581103,cite1_PhysRevE.72.026132}. 
Scholars have proposed methods based on local modularity \cite{cite4_article,cite25_ni2019community}, \textit{k}-core \cite{cite20_barbieri2015efficient}, \textit{k}-truss \cite{cite21_liu2020truss}, \textit{k}-clique\cite{cite22_cui2013online}, personalized PageRank \cite{cite23_yin2017local,cite24_kloumann2014community} and other methods.
Chen et al. \cite{cite2_5581103} identify local maximum-degree nodes that are adjacent to the seed nodes. They then merge communities that exhibit similarities above a predefined threshold to get the local communities associated with the seed nodes. 
Guo et al. \cite{cite4_article} divide the formation of local communities into two stages: the core area detection stage and the local community expansion stage. In the core area detection stage, they use local modularity density to ensure community quality. Subsequently, in the local community expansion phase, boundary nodes are identified based on the similarity between the nodes and the local community. 
Liu et al. \cite{cite21_liu2020truss} define the D-truss community search problem. To solve this problem, a local algorithm is designed to search for possible D-truss from the minimum query distance. 
Hao et al. \cite{cite23_yin2017local} convert the original network into a weighted graph using motifs, calculate approximate Personalized PageRank (PPR) vectors based on motifs for the neighborhood of seed nodes, and finally obtain the seed node's community through a sweep process.

The above studies focus on mining communities from a single network, ignoring that user activities often span multiple networks in the real world. 
Our work aims to mine the community structure by utilizing data from multiple networks for local community mining and using complementary information between multiple networks.
Therefore, the goal of our work is different from those of the above studies.
\subsection{Multiple Community Detection}
Users often participate in multiple networks simultaneously\cite{cite68_10414096}. 
Relying solely on data from a single network fails to provide a comprehensive understanding of users' behaviors\cite{cite72_kurant2006layered,cite73_buldyrev2010catastrophic,cite74lee2012correlated}. 
Since utilizing complementary information from diverse networks enhances community quality, multiple community detection methodologies have emerged, which are broadly classified into three categories
\cite{cite28_magnani2021community}.
The first category involves directly integrating multiple networks into a unified single network, applying traditional single network community detection methods to obtain the network's community structure \cite{cite10_YILDIRIMOGLU2018254,cite11_han2015consistent,cite12_rocklin2013clustering}. 
Yildirimoglu et al. \cite{cite10_YILDIRIMOGLU2018254} explore community structure in urban transportation networks. 
They construct Voronoi diagrams with multiple cells from public transportation trajectory networks, public transportation passenger trajectory networks, and road traffic trajectory networks. Subsequently, a modularity-maximizing approach is employed to determine the community structure.
The second category involves applying single network community detection methods to independently identify communities for each network. 
Subsequently, these individually detected communities are integrated to construct comprehensive  community structures
\cite{cite13_berlingerio2013abacus,cite14_dalibard2012community}. 
Michele et al. \cite{cite13_berlingerio2013abacus} design a multi-network community detection approach based on frequent pattern mining, which extracted frequent closed itemsets from community detected by label propagation algorithm \cite{cite47_raghavan2007near} from each single network to obtain multi-dimensional communities.
The third category involves integrating the information from each network to obtain the community structure on a multiple networks \cite{cite15_kumar2011co,cite16_liu2013multi,cite17_zong2018weighted,cite18_wang2019gmc}. 
Wang et al. \cite{cite18_wang2019gmc} construct a similarity matrix for each network and synthesize these into a composite fusion matrix. By jointly learning the fusion matrix and the similarity matrices, the final community structures are obtained.

The aforementioned studies aim to identify all communities within multiple networks, which need to utilize information from the entire multiple networks. In contrast, this paper concentrates on local community structure that includes a seed node within multiple networks, exploring only the neighborhood surrounding the seed node. 
Consequently, our work differs significantly from these previous approaches.
\subsection{Local Multiple Community Detection}

Considering that users are often interested in identifying communities containing specific nodes, some studies focus on mining such communities across multiple networks.
Roberto et al. \cite{cite50_interdonato2017local} propose a method called ML-LCD for mining community structures in multiple networks. 
ML-LCD provides the objective function according to the different ways to incorporate within-layer and across-layer topological features and use a greedy strategy to find the local communities on multiple networks. 
Ehsan et al. \cite{cite27_pournoor2021propagation} and Luo et al. \cite{cite6_10.1145/3394486.3403069}  explore local community structures using a random walk approach.
Specifically, Ehsan et al. \cite{cite27_pournoor2021propagation} consider both intra-layer and inter-layer probability to launch a random walker either to the neighboring nodes or nodes in other networks to obtain the community to which the seed node belongs. 
Luo et al. \cite{cite6_10.1145/3394486.3403069} deploy a random walker for each network. It considers the intra-layer probability of the current network and the transition probabilities from other networks. 
Li et al. design algorithms for multi-layer complex networks with direct influence/indirect influence relationships \cite{cite53_DBLP:journals/access/LiXLXJH19} and trust relationships \cite{cite54_DBLP:journals/winet/LiTTCY20}, respectively. The algorithm first accurately identifies the core nodes in the community, follows with a cluster analysis of these nodes, and ultimately obtains the local community based on these core nodes.

Existing studies on multi-view networks are inapplicable for mining communities within multi-domain networks. Furthermore, studies suited for multi-domain networks often neglect the consistency of communities across these networks and require access to the entire network, thereby consuming more resources than localized algorithms. Our study aims to optimize the quality of community structures within each network while ensuring their consistency across networks by examining only the local network areas around seed nodes. Therefore, our approach differs from prior studies.

\section{Conclusion} \label{conlusion}
In this paper, the LCDMN problem is investigated to leverage information from multiple networks to identify the community containing the seed node in each network, while ensuring that communities across different networks are as consistent as possible,  exploring only the local network around the seed node.
To address this problem, we propose a local community detection method with node affiliation, called LAMA, which uses local information. 
LAMA iteratively optimizes node affiliations and expands the community outward based on affiliations to detect community. 
LAMA optimizes node affiliations to enhance community quality within each network while maintaining consistent community structures across the networks.
Experiments conducted on two synthetic datasets and five real datasets demonstrate that LAMA  outperforms comparison methods.

\section*{Acknowledgments}
This work was supported by the National Natural Science Foundation of China [No.62106004 and No.62272001] and Guangdong Provincial Key Laboratory of Novel Security Intelligence Technologies [No.2022B1212010005].


\bibliographystyle{IEEEtran}
\bibliography{test}

\begin{thebibliography}{10}
\providecommand{\url}[1]{#1}
\csname url@samestyle\endcsname
\providecommand{\newblock}{\relax}
\providecommand{\bibinfo}[2]{#2}
\providecommand{\BIBentrySTDinterwordspacing}{\spaceskip=0pt\relax}
\providecommand{\BIBentryALTinterwordstretchfactor}{4}
\providecommand{\BIBentryALTinterwordspacing}{\spaceskip=\fontdimen2\font plus
\BIBentryALTinterwordstretchfactor\fontdimen3\font minus \fontdimen4\font\relax}
\providecommand{\BIBforeignlanguage}[2]{{%
\expandafter\ifx\csname l@#1\endcsname\relax
\typeout{** WARNING: IEEEtran.bst: No hyphenation pattern has been}%
\typeout{** loaded for the language `#1'. Using the pattern for}%
\typeout{** the default language instead.}%
\else
\language=\csname l@#1\endcsname
\fi
#2}}
\providecommand{\BIBdecl}{\relax}
\BIBdecl

\bibitem{cite2_5581103}
Q.~Chen and T.-T. Wu, ``A method for local community detection by finding maximal-degree nodes,'' in \emph{2010 International Conference on Machine Learning and Cybernetics}, Qingdao, China, 2010, pp. 8--13.

\bibitem{cite4_article}
K.~Guo, X.~Huang, L.~Wu, and Y.~Chen, ``Local community detection algorithm based on local modularity density,'' \emph{Applied Intelligence}, vol.~52, pp. 1238--1253, 2022.

\bibitem{cite25_ni2019community}
L.~Ni, P.~ManMan, J.~Wenjun, and L.~Kenli, ``A community detection algorithm based on multi-similarity method,'' \emph{Cluster Computing}, vol.~22, pp. 2865--2874, 2019.

\bibitem{cite44_luo2018local}
W.~Luo, D.~Zhang, H.~Jiang, L.~Ni, and Y.~Hu, ``Local community detection with the dynamic membership function,'' \emph{IEEE Transactions on Fuzzy Systems}, vol.~26, pp. 3136--3150, 2018.

\bibitem{cite23_yin2017local}
H.~Yin, A.~R. Benson, J.~Leskovec, and D.~F. Gleich, ``Local higher-order graph clustering,'' in \emph{Proceedings of the 23rd ACM SIGKDD international conference on knowledge discovery and data mining}, Halifax NS Canada, 2017, pp. 555--564.

\bibitem{cite52_DBLP:journals/tkde/NiGZLS24}
L.~Ni, J.~Ge, Y.~Zhang, W.~Luo, and V.~S. Sheng, ``Semi-supervised local community detection,'' \emph{{IEEE} Transactions on Knowledge and Data Engineering}, vol.~36, pp. 823--839, 2024.

\bibitem{cite63_10.1145/3394486.3403154}
Y.~Zhang, Y.~Xiong, Y.~Ye, T.~Liu, W.~Wang, Y.~Zhu, and P.~S. Yu, ``Seal: Learning heuristics for community detection with generative adversarial networks,'' in \emph{Proceedings of the 26th ACM SIGKDD International Conference on Knowledge Discovery \& Data Mining}, Virtual Event, CA, USA, 2020, p. 1103–1113.

\bibitem{cite64_10.1145/3336191.3371806}
D.~Luo, J.~Ni, S.~Wang, Y.~Bian, X.~Yu, and X.~Zhang, ``Deep multi-graph clustering via attentive cross-graph association,'' in \emph{Proceedings of the 13th International Conference on Web Search and Data Mining}, Houston, TX, USA, 2020, p. 393–401.

\bibitem{cite49_ni2018co}
J.~Ni, S.~Chang, X.~Liu, W.~Cheng, H.~Chen, D.~Xu, and X.~Zhang, ``Co-regularized deep multi-network embedding,'' in \emph{Proceedings of the 2018 world wide web conference}, Lyon, France, 2018, pp. 469--478.

\bibitem{cite60_9395530}
G.~Chao, S.~Sun, and J.~Bi, ``A survey on multiview clustering,'' \emph{IEEE transactions on artificial intelligence}, vol.~2, pp. 146--168, 2021.

\bibitem{cite61_DBLP:journals/bigdatama/YangW18}
Y.~Yang and H.~Wang, ``Multi-view clustering: {A} survey,'' \emph{Big Data Mining and Analytics}, vol.~1, pp. 83--107, 2018.

\bibitem{cite6_10.1145/3394486.3403069}
D.~Luo, Y.~Bian, Y.~Yan, X.~Liu, J.~Huan, and X.~Zhang, ``Local community detection in multiple networks,'' in \emph{Proceedings of the 26th ACM SIGKDD international conference on knowledge discovery \& data mining}, New York, NY, USA, 2020, pp. 266--274.

\bibitem{cite48_DBLP:journals/sigmod/KimL15}
J.~Kim and J.~Lee, ``Community detection in multi-layer graphs: {A} survey,'' \emph{ACM SIGMOD Record}, vol.~44, pp. 37--48, 2015.

\bibitem{cite46_li2018community}
X.~Li, G.~Xu, and M.~Tang, ``Community detection for multi-layer social network based on local random walk,'' \emph{Journal of Visual Communication and Image Representation}, vol.~57, pp. 91--98, 2018.

\bibitem{cite27_pournoor2021propagation}
E.~Pournoor, Z.~Mousavian, A.~Nowzari-Dalini, and A.~Masoudi-Nejad, ``A propagation-based seed-centric local community detection for multilayer environment: The case study of colon adenocarcinoma,'' \emph{Plos one}, vol.~16, p. e0255718, 2021.

\bibitem{cite66_9186335}
L.~Li and H.~He, ``Bipartite graph based multi-view clustering,'' \emph{IEEE Transactions on Knowledge and Data Engineering}, vol.~34, pp. 3111--3125, 2022.

\bibitem{cite71_DBLP:journals/axioms/KhanASN23}
M.~A. Khan, Adnan, T.~Saeed, and E.~R. Nwaeze, ``A new advanced class of convex functions with related results,'' \emph{Axioms}, vol.~12, p. 195, 2023.

\bibitem{cite28_magnani2021community}
M.~Magnani, O.~Hanteer, R.~Interdonato, L.~Rossi, and A.~Tagarelli, ``Community detection in multiplex networks,'' \emph{ACM Computing Surveys}, vol.~54, pp. 1--35, 2021.

\bibitem{cite16_liu2013multi}
J.~Liu, C.~Wang, J.~Gao, and J.~Han, ``Multi-view clustering via joint nonnegative matrix factorization,'' in \emph{Proceedings of the 2013 SIAM international conference on data mining}, Austin, Texas, {USA}, 2013, pp. 252--260.

\bibitem{cite30_greene2005producing}
D.~Greene and P.~Cunningham, ``Producing accurate interpretable clusters from high-dimensional data,'' in \emph{Proceedings of the 9th European Conference on European Conference on Machine Learning and Principles and Practice of Knowledge Discovery in Databases}, {Porto, Portugal}, 2005, pp. 486--494.

\bibitem{cite35_craven1998learning}
M.~Craven, D.~DiPasquo, D.~Freitag, A.~McCallum, T.~Mitchell, K.~Nigam, and S.~Slattery, ``Learning to extract symbolic knowledge from the world wide web,'' in \emph{Proceedings of the Fifteenth National/Tenth Conference on Artificial Intelligence/Innovative Applications of Artificial Intelligence}, {Madison, Wisconsin, USA}, 1998, p. 509–516.

\bibitem{cite50_interdonato2017local}
R.~Interdonato, A.~Tagarelli, D.~Ienco, A.~Sallaberry, and P.~Poncelet, ``Local community detection in multilayer networks,'' \emph{Data Mining and Knowledge Discovery}, vol.~31, pp. 1444--1479, 2017.

\bibitem{cite70_10414283}
L.~Zhang, B.~Li, L.~Ni, H.~Yang, and R.~Cao, ``Evolutionary multitasking local community detection on attributed networks,'' \emph{IEEE Transactions on Emerging Topics in Computational Intelligence}, vol.~8, pp. 1624--1639, 2024.

\bibitem{cite69_10.1145/1401890.1402008}
J.~Tang, J.~Zhang, L.~Yao, J.~Li, L.~Zhang, and Z.~Su, ``Arnetminer: extraction and mining of academic social networks,'' in \emph{Proceedings of the 14th ACM SIGKDD International Conference on Knowledge Discovery and Data Mining}, Las Vegas, Nevada, USA, 2008, p. 990–998.

\bibitem{cite1_PhysRevE.72.026132}
A.~Clauset, ``Finding local community structure in networks,'' \emph{Physical review E}, vol.~72, p. 026132, 2005.

\bibitem{cite20_barbieri2015efficient}
N.~Barbieri, F.~Bonchi, E.~Galimberti, and F.~Gullo, ``Efficient and effective community search,'' \emph{Data mining and knowledge discovery}, vol.~29, pp. 1406--1433, 2015.

\bibitem{cite21_liu2020truss}
Q.~Liu, M.~Zhao, X.~Huang, J.~Xu, and Y.~Gao, ``Truss-based community search over large directed graphs,'' in \emph{Proceedings of the 2020 ACM SIGMOD International Conference on Management of Data}, Portland OR USA, 2020, pp. 2183--2197.

\bibitem{cite22_cui2013online}
W.~Cui, Y.~Xiao, H.~Wang, Y.~Lu, and W.~Wang, ``Online search of overlapping communities,'' in \emph{Proceedings of the 2013 ACM SIGMOD international conference on Management of data}, New York, NY, USA, 2013, pp. 277--288.

\bibitem{cite24_kloumann2014community}
I.~M. Kloumann and J.~M. Kleinberg, ``Community membership identification from small seed sets,'' in \emph{Proceedings of the 20th ACM SIGKDD international conference on Knowledge discovery and data mining}, New York, NY, USA, 2014, pp. 1366--1375.

\bibitem{cite68_10414096}
Y.~Yu and D.~Sun, ``Incomplete multi-view clustering based on dynamic dimensionality reduction weighted graph learning,'' \emph{IEEE Access}, vol.~12, pp. 19\,087--19\,099, 2024.

\bibitem{cite72_kurant2006layered}
M.~Kurant and P.~Thiran, ``Layered complex networks,'' \emph{Physical review letters}, vol.~96, p. 138701, 2006.

\bibitem{cite73_buldyrev2010catastrophic}
S.~V. Buldyrev, R.~Parshani, G.~Paul, H.~E. Stanley, and S.~Havlin, ``Catastrophic cascade of failures in interdependent networks,'' \emph{Nature}, vol. 464, pp. 1025--1028, 2010.

\bibitem{cite74lee2012correlated}
K.-M. Lee, J.~Y. Kim, W.-k. Cho, K.-I. Goh, and I.~Kim, ``Correlated multiplexity and connectivity of multiplex random networks,'' \emph{New Journal of Physics}, vol.~14, p. 033027, 2012.

\bibitem{cite10_YILDIRIMOGLU2018254}
M.~Yildirimoglu and J.~Kim, ``Identification of communities in urban mobility networks using multi-layer graphs of network traffic,'' \emph{Transportation Research Part C: Emerging Technologies}, vol.~89, pp. 254--267, 2018.

\bibitem{cite11_han2015consistent}
Q.~Han, K.~Xu, and E.~Airoldi, ``Consistent estimation of dynamic and multi-layer block models,'' in \emph{International Conference on Machine Learning}, Lille, France, 2015, pp. 1511--1520.

\bibitem{cite12_rocklin2013clustering}
M.~Rocklin and A.~Pinar, ``On clustering on graphs with multiple edge types,'' \emph{Internet Mathematics}, vol.~9, pp. 82--112, 2013.

\bibitem{cite13_berlingerio2013abacus}
M.~Berlingerio, F.~Pinelli, and F.~Calabrese, ``Abacus: frequent pattern mining-based community discovery in multidimensional networks,'' \emph{Data Mining and Knowledge Discovery}, vol.~27, pp. 294--320, 2013.

\bibitem{cite14_dalibard2012community}
V.~Dalibard, ``Community detection in multi-layer networks,'' Ph.D. dissertation, Master’s thesis, University of Cambridge, 2012.

\bibitem{cite47_raghavan2007near}
U.~N. Raghavan, R.~Albert, and S.~Kumara, ``Near linear time algorithm to detect community structures in large-scale networks,'' \emph{Physical review E}, vol.~76, p. 036106, 2007.

\bibitem{cite15_kumar2011co}
A.~Kumar, P.~Rai, and H.~Daum\'{e}, ``Co-regularized multi-view spectral clustering,'' in \emph{Proceedings of the 24th International Conference on Neural Information Processing Systems}, Red Hook, NY, USA, 2011, p. 1413–1421.

\bibitem{cite17_zong2018weighted}
L.~Zong, X.~Zhang, X.~Liu, and H.~Yu, ``Weighted multi-view spectral clustering based on spectral perturbation,'' in \emph{Proceedings of the AAAI conference on artificial intelligence}, vol.~32, New Orleans, Louisiana, USA, 2018.

\bibitem{cite18_wang2019gmc}
H.~Wang, Y.~Yang, and B.~Liu, ``Gmc: Graph-based multi-view clustering,'' \emph{IEEE Transactions on Knowledge and Data Engineering}, vol.~32, pp. 1116--1129, 2019.

\bibitem{cite53_DBLP:journals/access/LiXLXJH19}
X.~Li, G.~Xu, W.~Lian, H.~Xian, L.~Jiao, and Y.~Huang, ``Multi-layer network local community detection based on influence relation,'' \emph{{IEEE} Access}, vol.~7, pp. 89\,051--89\,062, 2019.

\bibitem{cite54_DBLP:journals/winet/LiTTCY20}
X.~Li, Q.~Tian, M.~Tang, X.~Chen, and X.~Yang, ``Local community detection for multi-layer mobile network based on the trust relation,'' \emph{Wireless Networks}, vol.~26, pp. 5503--5515, 2020.

\end{thebibliography}

\begin{IEEEbiography}[{\includegraphics[width=1in,height=1.25in,clip,keepaspectratio]{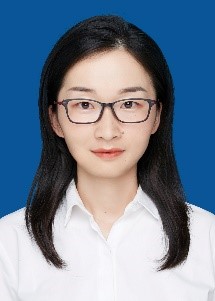}}]{Li Ni}
	received the PhD. degree from University of Science and Technology of China in 2020, and BE degree from Anhui University in 2015. 
	She is presently as a lecturer of School of Computer Science and Technology, Anhui University, Hefei, China. Her research interests include machine learning and data mining.
\end{IEEEbiography}
\vspace{-1 cm}
\begin{IEEEbiography}[{\includegraphics[width=1in,height=1.25in,clip,keepaspectratio]{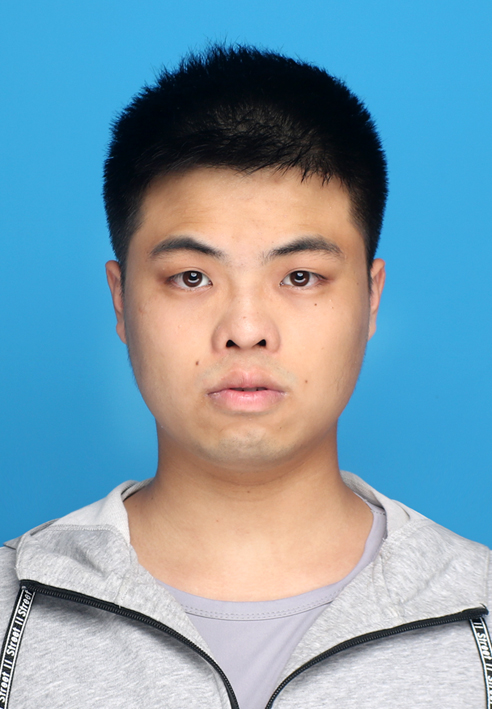}}]
{Zhou Xie}
received the B.S. degree from Anhui University of Science and Technology, China, in 2022.
Currently, he is working toward the MSc degree in the School of Computer Science and Technology, Anhui University, China.
His research interests include data mining.
\end{IEEEbiography}
\vspace{-1 cm}
\begin{IEEEbiography}[{\includegraphics[width=1in,height=1.25in,clip,keepaspectratio]{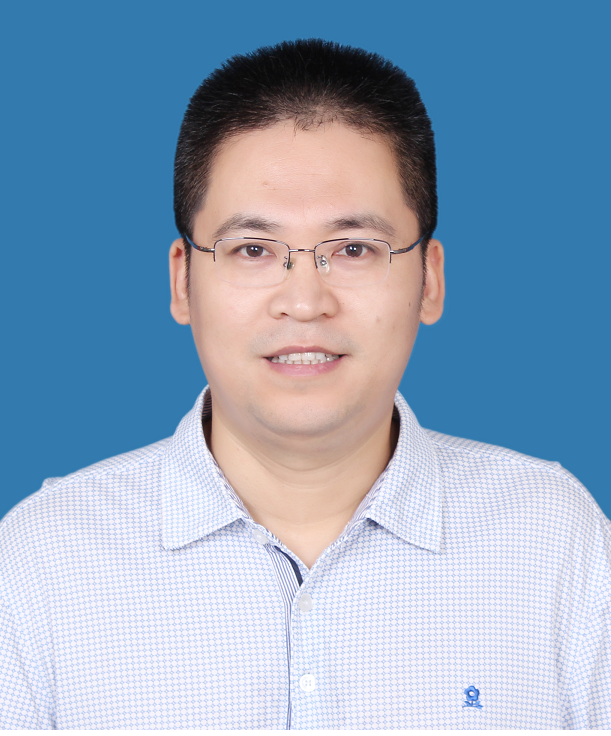}}]
{Yiwen Zhang}
received the Ph.D. degree in management science and engineering from the Hefei University of Technology, Anhui, China, in 2013.
He is a Professor with the School of Computer Science and Technology, Anhui University, China.
His current research interests include service computing, cloud computing, and big data.
\end{IEEEbiography}
\vspace{-1 cm}
\begin{IEEEbiography}[{\includegraphics[width=1in,height=1.25in,clip,keepaspectratio]{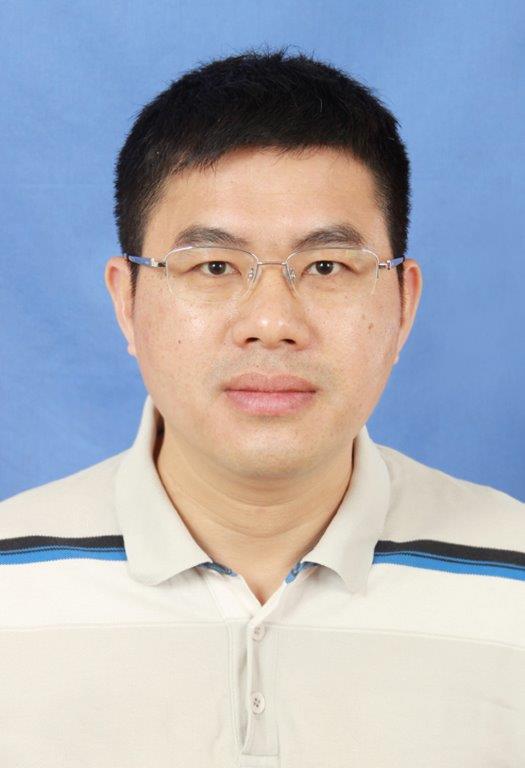}}]{Wenjian Luo}
	received the BS and PhD degrees from Department of Computer Science and Technology, University of Science and Technology of China, Hefei, China, in 1998 and 2003. He is presently a professor of School of Computer Science and Technology, Harbin Institute of Technology, Shenzhen, China. His current research interests include computational intelligence and applications, network security and data privacy, machine learning and data mining. He is a distinguished member of CCF and a senior member of IEEE, ACM and CAAI. He currently serves as an associate editor or editorial board member for several journals including Information Sciences Journal, Swarm and Evolutionary Computation Journal, Journal of Information Security and Applications, Applied Soft Computing Journal and Complex \& Intelligent Systems Journal. Currently he also serves as the chair of the IEEE CIS ECTC Task Force on Artificial Immune Systems. He has been a member of the organizational team of more than ten academic conferences, in various functions, such as program chair, symposium chair and publicity chair.
\end{IEEEbiography}
\vspace{-1.2 cm}
\begin{IEEEbiography}[{\includegraphics[width=1in,height=1.25in,clip,keepaspectratio]{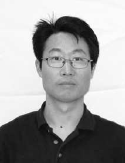}}]{Victor S.Sheng}
received the master's degree in computer science from the University of New Brunswick, Canada, in 2003 and the PhD degree in computer science from Western University, Ontario, Canada, in 2007. He is an associate professor of computer science with Texas Tech University. His research interests include data mining, machine learning, and related applications. He was an associate research scientist and NSERC postdoctoral follow in information systems in the Stern Business School, New York University. He received the test-of-time award from KDD'20, the best paper award runner-up from KDD'08, and the best paper award from ICDM '11. He is a PC member for a number of international conferences and a reviewer for several international journals. He is a senior member of the IEEE and a lifetime member of the ACM.
\end{IEEEbiography}

\vfill

\end{document}